\documentclass[a4paper,12pt]{article}
\pdfoutput=1
\usepackage{amssymb,amsmath,bm}
\usepackage{epigraph}
\usepackage{color}
\usepackage{slashed}
\usepackage{cite}
\usepackage[latin1]{inputenc}
\usepackage{amsfonts}
\usepackage{lscape}
\usepackage{amsthm}
\usepackage{booktabs}
\usepackage{array}
\usepackage{rotating}
\usepackage{float}
\usepackage{multirow}
\usepackage{graphicx,rotating,amsmath,multirow,xcolor,colortbl,xcolor}
\usepackage{hyperref,pdflscape,slashed}

\font\tenrsfs=rsfs10 at 12pt
\font\sevenrsfs=rsfs7
\font\fiversfs=rsfs5
\newfam\rsfsfam
\textfont\rsfsfam=\tenrsfs
\scriptfont\rsfsfam=\sevenrsfs
\scriptscriptfont\rsfsfam=\fiversfs
\def\mathscr#1{{\fam\rsfsfam\relax#1}}

\usepackage[small,bf]{caption}
\newcommand{\bea}{\begin{eqnarray}}
\newcommand{\eea}{\end{eqnarray}}
\newcommand{\beq}{\begin{equation}}
\newcommand{\eeq}{\end{equation}}

\begin{document}
\begin{center}
\boldmath

{\textbf{\Large On Future High-Energy Colliders}}

\unboldmath

\bigskip

\vspace{0.3 truecm}

{\bf Gian Francesco Giudice} 
 \\[3mm]

{\it CERN, Theoretical Physics Department, Geneva, Switzerland}\\[2mm]

\vspace{0.8cm}

{\bf Abstract }

\begin{quote}
An outline of the physics reasons to pursue a future programme in high-energy colliders is presented.
\end{quote}

\end{center}

\vspace{0.8cm}

While the LHC physics programme is still in full swing, the preparations for the European Strategy for Particle Physics and the recent release of the FCC Conceptual Design Report bring attention to the future of high-energy physics. How do results from the LHC impact the future of particle physics? Why are new high-energy colliders needed?

\subsubsection*{Where do we stand after the LHC discovery of the Higgs boson?}

Undoubtedly, the highlight of the LHC programme so far has been the discovery of the Higgs boson, which was announced at CERN on 4 July 2012. The result has had a profound impact on particle physics, establishing new fundamental questions and opening a new experimental programme aimed at exploring the nature of the newly found particle. The revolutionary aspect of this discovery can be understood by comparing it with the previous discovery of a new elementary particle -- the one of the top quark, which took place at Fermilab in 1995. The situation at the time was completely different. The properties of the top quark were exactly what was needed to complete satisfactorily the existing theoretical framework and the new particle fell into its place like the missing piece of a jigsaw puzzle. Instead, the discovery of the Higgs boson leaves us wondering about many questions still left unanswered. The top quark was the culmination of a discovery process; the Higgs boson appears to be the starting point of an exploration process.  

The Higgs boson is not simply another particle to be added to the list of the fundamental building blocks of matter, but it is the manifestation of a completely new physical phenomenon, whose dynamical origin still remains poorly understood and largely mysterious in the context of the Standard Model. This is the phenomenon of spontaneous symmetry breaking -- something that has been seen at play in superconductivity and other condensed-matter systems, but never before in the realm of elementary particles. The Higgs boson is something truly different from anything we have found so far in the particle world. Paradoxically, the intricacy of the Higgs boson lies in its simplicity. Unlike any other known elementary particle, the Higgs boson has no spin, no electric charge, and it superficially appears as a simple bundle of mass. This simplicity is at the origin of a puzzle that has mystified theoretical physicists for decades. 

For every massive particle, it is always possible to conceive an observer that catches up with the particle, thus seeing it at rest. When viewed at rest, a particle cannot distinguish any preferred direction, as a consequence of the rotational invariance of empty space. Therefore, a massive particle must be allowed to spin along all possible directions and must contain all possible polarisation states. This is not necessarily true for a massless particle which, according to special relativity, must travel at the speed of light and no physical observer can ever measure it at rest. This is why a massless particle can be found spinning clockwise with respect to the direction of motion, while lacking the state corresponding to anti-clockwise spin. In other words, for spinning particles there is a clear distinction between being massless or massive because the two cases correspond to a different number of quantum states. This is not true for a spinless particle like the Higgs boson, for which the massive and massless cases are described by exactly the same number of quantum states. 

The conundrum shows up when quantum mechanics enters the game. One of the rules of the uncanny world of quantum mechanics is that everything not forbidden is compulsory. This rule is due to the ubiquitous quantum fluctuations which blur every physical system, rendering special and non-generic configurations highly unlikely. For spinning particles, the mismatch of states between the massless and massive cases forbids quantum fluctuations to pump energy into a massless particle and turn it into a massive one. But nothing prevents them from doing so in the case of a spinless particle and indeed theoretical calculations predict that the Higgs boson should gain an enormous mass in any realistic quantum system. The discovery of a relatively light Higgs boson clashes with the logic of quantum mechanics, leaving theoretical physicists bewildered. This problem is known as `Higgs naturalness.' 

Another important consequence of the structural simplicity of the Higgs boson is that it provides a natural bridge between the known particle world -- the one described by the Standard Model -- and other possible hidden sectors, related perhaps to dark matter or to other still-undiscovered particles. This special property is related, once again, to the spinless nature of the Higgs boson. Within the Standard Model, the Higgs boson is the only particle that can form combinations that preserve all spacetime symmetries and interact with hidden particle sectors in a way that is not suppressed by short-distance scales. This unique property makes the Higgs boson one of the best keys at our disposal to unlock the door towards the mysteries of possible hidden worlds.

One of the most striking results of the LHC exploration was the discovery of a completely new type of force.
Before the discovery of the Higgs boson, we knew of four fundamental forces governing natural phenomena (strong, weak, electromagnetic forces and gravity). All of them were successfully deduced from a single conceptual principle -- called `gauge symmetry.' In the meantime, the LHC has discovered the existence of new forces in nature. These forces control how the Higgs boson interacts with quarks and leptons and, so far, the LHC has identified their effect on top and bottom quarks and on tau leptons (the so-called `third generation').  According to the Standard Model, these new forces are as fundamental as those previously known, but of a different nature -- not `gauge-like.' While the Standard Model properly describes these forces, it is silent on the origin of the many free parameters associated with them or on any deeper explanation of their complex structure. 

The `non-gauge' nature of the newly discovered forces means that we are confronted with a completely new phenomenon. For instance, the interaction strengths of the new forces are not quantised, unlike all other quantum forces we have probed so far. When compared with the logical purity and conceptual simplicity of the `gauge forces,' the new forces associated with the Higgs boson look provisional. Their structure raises conceptual puzzles, which are not necessarily inconsistencies of the Standard Model, but possible clues indicating the existence of a deeper theoretical description.  

While the Higgs boson is playing a central role in studies at the LHC, it is only one aspect of a very broad scientific programme. The LHC has already produced a wealth of precision measurements on parameters as fundamental as the $W$ and top-quark masses, on important properties of $B$ mesons, and on observables relevant to strong interactions. With these measurements we are acquiring refined knowledge about the Standard Model, which is a necessary prerequisite to advance our understanding of the particle world and to search for new phenomena.

\subsubsection*{The status of high-energy physics}
At the end of the 19$^{\rm th}$  century, physics was in a satisfactory state, able to describe all known phenomena. This was summarised by the slogan: ``There is nothing more to discover in physics; all that remains is more and more precise measurements." Nonetheless, in a famous speech delivered in 1900, Lord Kelvin identified a few `clouds' obscuring the bright sky of 19$^{\rm th}$  century physics. Luckily, some physicists did not ignore those clues: relativity and quantum mechanics followed.

Similarly, today the Standard Model of particle physics, together with the current model of cosmology and a sufficient number of free parameters, are able to describe all known physical phenomena and the large-scale properties of the universe. But, watching attentively, we see some `clouds' in the horizon. 

Puzzles in today's particle physics come from structural problems of the Standard Model (the nature of the Higgs boson, Higgs naturalness, the origin of symmetry breaking dynamics, the stability of the Higgs potential, the existence of three generations of matter, the pattern of quark and lepton masses and mixings, the dynamics generating neutrino masses), from embedding the Standard Model into a broader framework (the unification of forces, quantum gravity, the cosmological constant), from attempts to give particle-physics explanations of cosmological properties (the nature and origin of dark matter, dark energy, cosmic baryon asymmetry, inflation). 

Just as at the end of the 19$^{\rm th}$  century, one may argue today that the Standard Model is satisfactory and there is nothing more to discover in particle physics, other than making more precise measurements. Nonetheless, clues should not be dismissed because those `clouds' could be symptoms of a more fundamental disease. The real limitation of 19$^{\rm th}$  century physics was describing nature in terms of disconnected theories (mechanics, gravitation, electromagnetism, thermodynamics, optics, etc.) lacking a unified vision. Today, the Standard Model's shortcomings may be indicators of a deep structural limitation of our theoretical description. Without searching, we will never find answers.

\subsubsection*{Special puzzles for the LHC}
Certainly not all the problems listed above can be addressed by the LHC, although studies at present and future colliders are essential to guide research. But out of the many `clouds' that could turn into storms shattering the Standard Model, two of them are playing a special role at the LHC: Higgs naturalness and dark matter. This is because there are good arguments to link these problems to the energy domain explored by the LHC. In other words, they are the most likely place to find surprises at the LHC. 

Starting from these clues, in the past decades theorists have elaborated hypotheses to address some of the shortcomings of the Standard Model, proposing new bold ideas such as supersymmetry, technicolour, extra dimensions, Higgs compositeness, and a variety of other speculative theories. So far, the LHC has given no positive indication for the existence of new phenomena, ruling out many versions of the postulated theories. 

This result is nothing but the scientific method at work. Experiments act as a `natural selection' process in which some theoretical hypotheses become extinct and others evolve, according to `survival of the fittest.' 
Although the LHC programme is still at an intermediate phase and it is premature to draw definitive conclusions, 
it is already clear that the LHC is radically reshaping our perspective on the particle world. The LHC experimental results are
forcing theorists to think differently about problems and to search for new solutions. In this situation of renewed theoretical exploration, experimental physics is needed -- more than ever -- to break new ground.

\subsubsection*{Measurements versus discoveries}
The rugged and twisted path towards scientific knowledge is punctuated not only by discoveries, but also by disproved hypotheses and null results that redirect research. As shown repeatedly by history, the non-discovery of expected results can be as effective as the discovery of unexpected results in igniting momentous paradigm changes. 

When 19$^{\rm th}$ century theorists were puzzled about how electromagnetic waves could propagate, they addressed the conundrum using the known framework of Maxwell's theory and added the hypothesis of a new medium permeating space: the aether. The lack of discoveries in the Michelson-Morley experiment ruled out this hypothesis and eventually triggered a much more revolutionary solution: special relativity.

When Le Verrier was puzzled by the discrepancy in the precession of Mercury's perihelion, he turned to Newton's gravity for a solution and predicted the existence of a new planet, named Vulcan. The planet was never discovered and the real solution turned out to be much more revolutionary: general relativity.

These two examples illustrate an identical pattern. When scientists are confronted with a problem, they first look for solutions within the accepted theoretical framework. When experiments declare a non-discovery, then scientists are forced to think differently and this may spark a paradigm change.

The lesson for us could be that the solutions devised so far by particle physicists to deal with the shortcomings of the Standard Model are overly rooted in the currently accepted framework. In spite of their bold appearance, the proposed theories only look like logical extensions of the same principles that govern the Standard Model. The LHC experimental results may push theoretical thinking into the territory of radical changes.

Measurements, and not only new discoveries, can provide the decisive steps for advancements in science and knowledge. A pertinent example is the story of LEP.  In spite of the lack of any discovery of new particles, LEP transformed particle physics like few other experimental projects before. It established the gauge paradigm as the ruling principle of the Standard Model, it pruned away a variety of alternative hypotheses, it substantiated the existence of only three generations of quarks and leptons (by `not discovering' additional light neutrinos), it brought the question of force unification to a quantitative level, it revolutionised the use of precision measurements to gain knowledge of short-distance phenomena. It is largely because of LEP that today we recognise the formidable power of gauge theories and the conceptual synthesis of the Standard Model.

\subsubsection*{What can we expect from future high-energy colliders?}
Fundamental research beyond the frontiers of knowledge is -- by definition -- unpredictable.
When Galileo perfected the telescope, he could not foretell how many moons were to be discovered around Jupiter. When we propose a new high-energy collider, we cannot predict the number of particles we are going to discover, but we can define the questions we want to address. The value of a scientific project dedicated to the exploration of the unknown does not lie in the number of foreseen new discoveries, but in the knowledge that will be gained from its results. 

With the information that we have gathered so far from the LHC, we can already identify some broad research programmes that have guaranteed goals of expanding our knowledge. Moreover, we know that only a vigorous programme of exploration can address the many fundamental questions that still remain unanswered in particle physics.
\medskip

\noindent  \textbf{\textit {The Higgs programme}}

The discovery of the Higgs boson has opened a new compelling experimental programme, whose goal is the precise determination of the new particle's properties. This programme is a search into unexplored territory because the dynamics behind the Higgs boson has still to be understood and the LHC has so far only scratched the surface of the phenomenon. Exploring the characteristics of the Higgs boson and its associated forces, establishing their nature and understanding their origin are essential objectives of future collider projects. Indeed, all existing proposals for future accelerators have the Higgs programme as a primary goal. 

The Higgs programme is a new frontier for particle physics, whose targets are well defined. The first goal is to measure the Higgs coupling constants (which determine the interaction strengths of all forces involving Higgs bosons) with precision down into the per-mille region. Such measurements are sensitive to any substructure possibly present inside the Higgs boson, if the particle is not truly elementary. Future collider projects can measure the inner Higgs structure with an astounding experimental resolution, which can reach distances $10^{5}$ times smaller than the size of a proton. The new Higgs forces have been measured so far only for the third generation of matter (top, bottom, tau). The next challenge is to test the Higgs forces for second-generation particles, especially the charm quark and the muon. Since the origin for the wide difference in mass between particle generations is still a mystery, probing Higgs interactions with second-generation particles is a conceptually new test of the mechanism that feeds mass into elementary particles. Another important test is the measurement of the so-called `invisible Higgs decay', which corresponds to the disintegration of the Higgs boson into particles that cannot be directly revealed by particle detectors at high-energy colliders, such as neutrinos. `Invisible Higgs decays' are especially interesting because they explore possible new types of elusive particles, maybe related to the nature of dark matter. Another target of the Higgs programme is the measurement of the Higgs self-interaction, which is a direct test of the new kind of force that is thought to be responsible for generating the non-trivial structure of the Higgs vacuum. Such a measurement will also provide us with the elements to reconstruct theoretically the details of the phase transition that is believed to have occurred in the universe only $10^{-11}$ seconds after the Big Bang. Finally, another goal is to measure a variety of rare Higgs decays, which are rich with important information. For instance, the Higgs decay into $Z$ and a photon is an efficient probe of new physics effects occurring at very short distances; Higgs decays into two different leptons ($\tau \mu$, $\tau e$ or $\mu e$) or into CP-violating final states can probe the nature of symmetries in the particle world.

Interestingly, the different proposals for future high-energy collider projects have a complementary role in the exploration of the Higgs properties. Linear or circular electron-positron colliders can measure the main Higgs interactions with high accuracy and make the first measurements of the as-yet untested interaction with the charm quark. Exploiting the formidable rates of Higgs-boson production, future proton-proton colliders can then pursue further these studies and test rare Higgs decays, including the measurement of the Higgs interaction with muons at the percent level. Moreover, future high-energy proton-proton colliders can measure with precision better than ten percent the Higgs self-interaction. Measurements at proton-proton colliders are most precise for ratios of Higgs couplings, where the large systematic uncertainties cancel out. For this reason, global studies benefit from combining results from hadron colliders with results from lepton colliders, whose cleaner environment allows for very precise absolute determinations of certain Higgs couplings. For instance, the synergy between different projects is illustrated by the measurement of the Higgs interaction with top quarks, which can reach the percent precision at proton-proton colliders only thanks to information derived from future electron-positron collider data.

These measurements are not done purely for the sake of precision, but because they provide us with new knowledge about the least understood and most puzzling sector of the Standard Model. From these measurements we will learn about untested fundamental forces of nature, about the dynamics of the space-time vacuum, and about the mechanism responsible for the masses of Standard Model particles. Our knowledge of the electroweak symmetry breaking mechanism will be radically different after a precision Higgs programme. Much as LEP transformed our knowledge of the Standard Model gauge sector, future colliders will transform our knowledge of the Higgs sector.

The goals of the Higgs programme may sound rather abstract. But, on the contrary, the Higgs boson is an essential element that determines the properties of the ordinary matter we are made of. This can be understood by imagining a hypothetical world in which one could dial the vacuum configuration of the Higgs field and change its intensity with respect to the value we observe in our world. By reducing the intensity of the Higgs field, one would gradually make atoms grow in size. All matter (including ourselves) would puff up, until bound atoms could no longer exist. Were one to rotate the dial in the opposite direction and make the Higgs field more intense, the consequences would be equally dramatic. If the vacuum configuration of the Higgs field were only a factor of five stronger than what we observe in our world, atomic nuclei would not be stable because neutrons would rapidly decay: hydrogen would be the only chemical element that could possibly exist in the universe. This would make for a pretty dull universe without any chemical structure -- let alone life. In summary, the universe in which we live depends critically on the existence of the Higgs field and matter would behave very differently if the Higgs boson differed even slightly from what we observe. Understanding the properties of the Higgs boson means understanding the underlying reasons for the observed structure of matter.

\medskip

\noindent  \textbf{\textit {The precision programme}}

One of the most important legacies of the LHC is about the experimental capabilities of hadron colliders. Progress in data analysis and detector performance, combined with advanced theoretical calculations, led to previously unimaginable precision in measurements at proton-proton colliders. Today the LHC is performing measurements that would have been unthinkable at the time in which the project was designed. This patrimony of experience opens up the possibility of a full programme of precision studies at future colliders. 

A campaign of precise measurements of all Standard Model observables is essential to test the theoretical framework and to sharpen the tools necessary to look for the small departures between theoretical predictions and experimental data that would indicate the presence of new phenomena. A crucial aspect of the precision programme rests on a characteristic feature of quantum mechanics. The result of a high-energy scattering process is sensitive not only to the properties of the particles that participate in the collision, but also to particles with masses much larger than the available initial energy. This counterintuitive result is due to quantum fluctuations, which bring to existence, even if only for a very short time, `virtual particles' whose presence naively seems to violate the requisite of energy conservation. These particles, although too heavy to be directly produced with the available energy of the collider, nonetheless leave an `indirect' trace in the scattering process as a consequence of their ephemeral `virtual' quantum existence. These `indirect' traces can be detected through a combined effort involving painstakingly precise experimental measurements and extremely accurate theoretical calculations of the Standard Model prediction. The quantum effects from virtual particles make precision measurements in high-energy collisions especially interesting because they allow us to probe, in specific cases, new phenomena occurring at distances much smaller than those directly explored by the collider energy. 

Another important aspect of `indirect' searches for new phenomena is that `virtual' effects generally grow with the energy of the collision. This means that the higher the collider energy, the more effective indirect searches become. Indirect searches are optimised by the right compromise between energy and precision. Different high-energy collider proposals have different strengths for reaching the best conditions. 

As for the Higgs programme, electron-positron and proton-proton colliders can complement each other in the precision programme. By measuring precisely the properties of known particles (such as the $W$, $Z$, and top quark) at extreme high-energy conditions, not only do we gain knowledge about those particles, but we also explore new phenomena that could help us to resolve some of the Standard Model's shortcomings. The stupendous number of $W$, $Z$, and top quarks that can be produced at future colliders provides an unprecedented probe of the detailed properties of the Standard Model and a powerful exploration tool into new phenomena.

In many respects, the Higgs and precision programmes are two aspects of the same scientific strategy. According to the Standard Model, the Higgs boson and the longitudinal polarisations of the $W$ and $Z$ bosons are only different states of the same fundamental entity. Therefore, precision Higgs and electroweak measurements are intimately related and complement each other in terms of the information that can be extracted from data.

A particularly interesting aspect of the precision programme is the study of a special class of decay processes involving quarks and leptons (the so-called `flavour physics'). These processes offer a unique opportunity for new discoveries because in the Standard Model, for accidental reasons related to the structure of the theory, they happen to be very rare or even absolutely forbidden. To a certain extent, the study of these processes investigates, from a different angle, the nature of the same new forces explored by the Higgs programme.

\medskip

\noindent  \textbf{\textit {The exploration programme}}

Exploration of the unknown is the main driver of fundamental science, and it embodies one of the aspirations that define human civilisation. The thirst for pure knowledge is a powerful vehicle for progress, which propagates from basic science into technological and social advancements. Particle physics has always been driven by the spirit of exploration, leading it to push the frontier of knowledge deep into the smallest fragments of spacetime and to unravel the natural phenomena that occur at the most minute distance scales. This exploration was rewarded with amazing insights into the fundamental laws of nature. The same spirit of exploration remains today the primary motivation for any high-energy collider project beyond the LHC. Exploration is the very spirit and essence of research. 

Addressing many of the mysteries in particle physics requires a bold step in the exploration of Nature at the smallest possible distances. With present available technology, this can be done only by means of high-energy colliders, which allow for direct observation of the phenomena that occur in the microworld. Breaking the frontier of knowledge is what drives particle physics towards exploration into ever smaller distances with more powerful colliders.

Many of the conceptual tools that led to progress in our understanding of the early evolution of the universe and its large-scale structure came -- almost paradoxically -- from our study of the microworld. The deep connection existing in nature between inner space and outer space first emerged with the understanding of how stars shine in terms of nuclear reactions and later led to the successful prediction of the cosmic abundance of light chemical elements in terms of nucleosynthesis. A more modern and pertinent example of the link between elementary particles and the cosmos is provided by inflation. According to this theory, a Higgs-like field is the engine driving the early evolution of the universe and its quantum fluctuations are the seeds that created the structures in matter and radiation that we observe in the sky.

Today particle physics and cosmology are inextricably intertwined. More generally, research in particle physics is evolving towards an interplay of different subjects addressing, from different angles, related questions in fundamental physics. In this situation it is futile to expect substantial progress by pushing only a single line of research. Further advancements require a global scientific effort with a diversified experimental programme that ranges from particle physics to observational cosmology, astroparticle physics and beyond. 

In the context of this broad scientific effort, high-energy colliders remain an indispensable and irreplaceable tool to continue our exploration of the inner workings of the universe. While each experimental technique can contribute to this search from a different and complementary perspective, high-energy colliders remain the best microscopes at our disposal, with a formidable exploration power into the mysteries of matter at short distances. No other instrument or research programme can replace high-energy colliders in the search for the fundamental laws governing the universe. 

\bigskip

\noindent{\bf Acknowledgements}

I am much indebted to my colleagues Fabiola Gianotti, Michelangelo Mangano, Matthew McCullough, Riccardo Rattazzi, and Gavin Salam for interesting discussions, useful suggestions, and for sharing with me their insight on the subject.

\end{document}